\documentclass[twocolumn,showpacs,preprintnumbers,amsmath,amssymb, superscriptaddress, prl, reprint, citeautoscript]{revtex4-1}

\usepackage{stmaryrd}
\usepackage{txfonts}
\usepackage{amssymb}
\usepackage{mathrsfs}
\usepackage{graphicx}% Include figure files
\usepackage{dcolumn}% Align table columns on decimal point
\usepackage{bm}% bold math
\usepackage{epsfig}
\usepackage{color}                    % For creating coloured text and background
\usepackage{hyperref}                 % For creating hyperlinks in cross references

\begin{document}
\preprint{\href{http://dx.doi.org/10.1103/PhysRevB.89.220405}{S. -Z. Lin \emph{et al.}, Phys. Rev. B {\bf{89}}, 220405(R) (2014).}}

\title{Magnetic field induced phases in anisotropic triangular antiferromagnets: application to $\mathrm{CuCrO_2}$}

\author{Shi-Zeng Lin}
\affiliation{National High Magnetic Field Laboratory (NHMFL), MPA-CMMS Group, Los Alamos National Laboratory, Los Alamos, New Mexico 87545, USA}
\affiliation{Theoretical Division, Los Alamos National Lab, Los Alamos, NM 87545, USA}

\author{Kipton Barros}
\affiliation{Theoretical Division, Los Alamos National Lab, Los Alamos, NM 87545, USA}

\author{Eundeok Mun}
\altaffiliation{Present address: Ames Laboratory, Iowa, USA}
\affiliation{National High Magnetic Field Laboratory (NHMFL), MPA-CMMS Group, Los Alamos National Laboratory, Los Alamos, New Mexico 87545, USA}

\author{Jae-Wook Kim}
\affiliation{National High Magnetic Field Laboratory (NHMFL), MPA-CMMS Group, Los Alamos National Laboratory, Los Alamos, New Mexico 87545, USA}

\author{Matthias Frontzek}
\affiliation{Laboratory for Neutron Scattering, Paul Scherrer Institute, CH-5232 Villigen, Switzerland}

\author{S. Barilo}
\affiliation{Institute of Solid State and Semiconductor Physics, Minsk 220 072, Belarus}

\author{S. V. Shiryaev}
\affiliation{Institute of Solid State and Semiconductor Physics, Minsk 220 072, Belarus}

\author{Vivien S. Zapf}
\affiliation{National High Magnetic Field Laboratory (NHMFL), MPA-CMMS Group, Los Alamos National Laboratory, Los Alamos, New Mexico 87545, USA}

\author{Cristian D. Batista}
\affiliation{Theoretical Division, Los Alamos National Lab, Los Alamos, NM 87545, USA}

%\author{Shi-Zeng Lin,$^{1,2}$, Kipton Barros,$^{1,2}$ Eundeok Mun,$^{2,*}$ Jae-Wook Kim,$^2$ Matthias Frontzek,$^3$ S. Barilo,$^4$ S. V. Shiryaev,$^4$ Vivien S. Zapf,$^2$ and Cristian D. Batista$^1$}

%\affiliation{$^1$Theoretical Division, Los Alamos National Lab (LANL), Los Alamos, NM 87545, USA}
%\affiliation{$^2$National High Magnetic Field Laboratory (NHMFL), MPA-CMMS group, LANL}
%\affiliation{$^3$Laboratory for Neutron Scattering, Paul Scherrer Institute, CH-5232 Villigen, Switzerland}
%\affiliation{$^4$Institute of Solid State and Semiconductor Physics, Minsk 220 072, Belarus}
%\altaffiliation{$^*$Now at Ames Laboratory, Iowa}

\begin{abstract}
We introduce a minimal spin model for describing the magnetic properties of $\mathrm{CuCrO_2}$. Our Monte Carlo simulations of this model reveal a rich magnetic field induced phase diagram, which explains  the measured  field dependence of the electric polarization. The sequence of phase transitions between different mutiferroic states arises from a subtle interplay between  spatial and spin anisotropy, magnetic frustration and  thermal fluctuations. Our calculations are compared to new measurements up to 92 T.
% which demonstrates how competing interactions can lead to rich magneto-electric functionality.
\end{abstract}
 \pacs{75.85.+t, 75.30.Kz, 77.80.-e, 75.50.Ee} %checked for Magnetic field induced phases in anisotropic triangular antiferromagnets: application to $\mathrm{CuCrO_2}$
\date{\today}
\maketitle

Triangular lattice antiferromagnets (TLA) are widely studied in the field of frustrated magnetism. Complex orderings and rich phase diagrams arise because three antiferromagnetic interactions within a triangle cannot be simultaneously satisfied. Delafossite CuCrO$_2$ is a particularly clean example of a TLA where quasi-classical Cr$^{3+}$ $S = 3/2$ spins form a  triangular lattice in the $ab$ plane. \cite{Kadowaki90,Crottaz1996} The spins have out-of-plane anisotropy and weak interlayer coupling  that is one to two orders of magnitude smaller than the in-plane interactions. \cite{Poienar10,Frontzek11,Vasiliev14}  The three spins of each triangle form a nearly 120$^\circ$ structure and all three sublattices form proper-screw spirals that propagate along the same [110] axis with  propagation vector  $q = 0.329$ (the spins rotate in and out of the $ab$ plane). ~\cite{Kadowaki90,Soda09,Soda10} The spiral can propagate along any of six directions (three choices for the [110] axis and two choices for the helicity) leading to six possible domains. 

The proper-screw spiral induces an electric polarization ${\bf P}$ along the spiral propagation vector.~\cite{Kimura08,Seki08,Kimura09PRL,Soda09,Poienar09,Poienar10,Soda10,Kajimoto10} This allows us to probe phase transitions between spiral states at high applied magnetic fields $\mathbf{H}_a$, while the magnetization is largely insensitive to these transitions. The non-zero  ${\bf P}$ is consistent with  Arima's mechanism for multiferroic behavior, ~\cite{Arima07} where the spiral magnetic structure slightly influences the hybridization between the Cr $d$-orbitals and the O $p$-orbitals via spin-orbit coupling, creating a net ${\bf P} \parallel {\bf q}$. Thus, a pattern of electromagnetic domains forms below the magnetic ordering temperature that can be influenced by small electric and magnetic fields relative to the dominant exchange interactions. \cite{Kimura09PRL,Soda09} 

\begin{figure}[t]
\psfig{figure=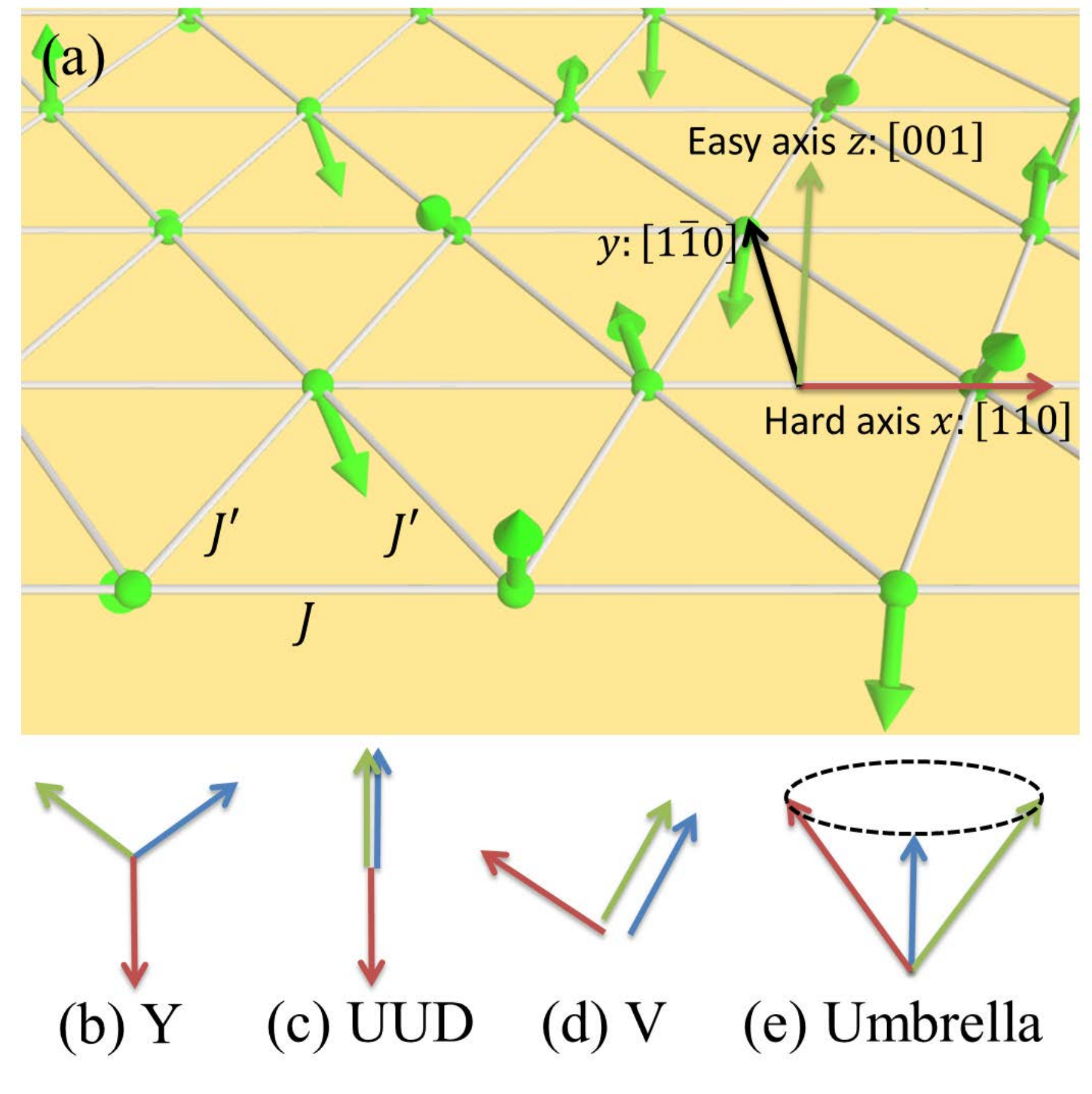,width=8.5cm}
\caption{(color online) (a) Schematic view of the Hamiltonian in Eq. \eqref{eq1}. (b-e) Typical spin configurations in a unit triangle. The Y state can be either commensurate (CY) or incommensurate with the lattice (ICY), while the umbrella state is always incommensurate (ICU) for $|J-J'|\ll J$. The UUD and V states are always commensurate with the lattice.
} \label{f1}
\end{figure}

The triangular layers of CuCrO$_2$ stack along the $c$-axis such that a Cr$^{3+}$ ion from one layer lies at the center of a triangle of Cr$^{3+}$ ions in the next layer.~\cite{Kadowaki90,Crottaz1996,Seki08} The triangular lattice distorts by about 0.01\% as a result of the spiral magnetic ordering, leading to two different exchange interactions, $J$ and $J'$, along different bonds of the triangle~\cite{Kimura09,Kimura09PRL,Poienar09,Poienar10,Aktas13} (Fig.~\ref{f1}). Thermodynamic measurements show two close-lying phase transitions. Elastic neutron diffraction measurements suggest that below $T_N = 24.2$ K, the triangular plane develops collinear spin correlations. A spiral long-range order appears below $T_{\mathrm{MF}} = 23.6$ K and also induces net $\mathbf{P}$, possibly via a first-order transition.  \cite{Kimura08,Ehlers13,Aktas13}

The ${\bf H}_a$ dependence  of this spiral ordering is only partially explored in experiments and theory.~\cite{Fishman11,Kimura08,Seki08,Mun13} For applied magnetic fields  along [110] and $H_a>5.3$ T, the proper screw spiral flops into a cycloidal spiral with the same ${\bf q}$ vector. ~\cite{Kimura08,Soda09,Kimura09PRL,Soda10,Yamaguchi10} Since there are six possible domains with different spiral propagation axes, the flop  only occurs in the two domains that have their propagation axis perpendicular to the applied magnetic field. During the spin flop, the electric polarization of those domains rotates from being perpendicular to being parallel to ${\bf H}_a$. This cycloidal spiral phase persists beyond 65 T.~\cite{Mun13} While the phase diagram for  ${\bf H}_a \parallel ab$ contains only one phase transition at 5.3 T (in the explored region of phase space up to 65 T), the phase diagram for ${\bf H}_a \parallel {\hat c}$ contains a series of field-induced phases.~\cite{Mun13} For certain temperatures, the sequence of phase transitions leads to an oscillation in the magnitude of  ${\bf P}$ as a function of $H_a$.~\cite{Mun13} 

Because  ${\bf P}$ is induced by a magnetic spiral in $\mathrm{CuCrO_2}$, it is interesting to know the magnetic structure of the new phases. We note that these phases are not captured by recent  calculations for $\mathrm{CuCrO_2}$.~\cite{Fishman11} Here we present a minimal model  that applies to $\mathrm{CuCrO_2}$, along with new measurements of ${\bf P}$ in CuCrO$_2$ up to 92 T. Our Monte Carlo (MC) simulations reproduce the zero-field spiral magnetic order and capture the essentials of the field-induced phase diagrams  along different magnetic field directions. Four key competing ingredients are important in this problem: frustration, thermal fluctuations, spatially anisotropic exchange interactions and spin anisotropy.  Although the spin anisotropy and spatial distortion are weak, they are always relevant perturbations because the ground state of the frustrated Heisenberg model is highly degenerate.

To build a minimal model for $\mathrm{CuCrO_2}$ we note  that the Cr$^{3+}$ $S=3/2$ spins are large enough to be treated classically, and that the new phases found in Ref. \cite{Mun13} occur at relatively high temperature $T$. In addition,  the low field spiral plane is perpendicular to $[110]$ and the electric polarization flop transition depends weakly on the in-plane field direction, indicating a weak in-plane hard-axis anisotropy. Finally, the ordered moment is maximal along $[001]$, implying that this is the easy-axis.~\cite{Frontzek11,Frontzek12} 
Based on these facts and the small spatial anisotropy (see Fig.~\ref{f1}), we introduce the following 2D model Hamiltonian for $\mathrm{CuCrO_2}$
\begin{equation}\label{eq1}
\mathcal{H}=\sum_{<ij>}J_{{ij}} \mathbf{S}_i\cdot \mathbf{S}_j+\sum _i\left[\frac{1}{2} A_x S_{i,x}^2-\frac{1}{2} A_z S_{i,z}^2-\mathbf{H}\cdot \mathbf{S}_i\right],
\end{equation}
where ${\bf H} \equiv g \mu_B {\bf H}_a$ ($\mu_B$ is Bohr magneton and $g \simeq 2$ is the $g$-factor),  $J$ and $J'$ are the AFM nearest neighbor (NN) interaction. The single-ion anisotropy terms are much weaker than the dominant exchange interactions, $0<A_x,\ A_z\ll J_{ij}$, and ${\bf S}_j$ is a classical unit vector representing the spin at site $j$. \footnote{A similar Hamiltonian with next NN AFM interaction and next-next NN AFM interaction was proposed based on  inelastic neutron scattering measurements.~\cite{Poienar10,Frontzek11} The parameters derived in Ref.~\onlinecite{Poienar10} however lead to a collinear ground state at zero magnetic field, which is inconsistent with experiments (see the supplemental information for details). In Ref.~\cite{Frontzek11}, the incommensurate spiral is stabilized by a ferromagnetic interlayer coupling because the layers are not vertically stacked along the $z$-axis. This Hamiltonian was used in Refs.~\cite{Fishman_Monte2012,Haraldsen_Spin2012,Fishman11} for $\mathrm{CuFeO_2}$ and $\mathrm{CuCrO_2}$. It is a subtle issue whether the incommensurability results from the inequivalent intra-layer bonds or/and from the weak frustrated interlayer coupling.  Here we seek for a minimal 2D model with only anisotropic NN AFM interactions to qualitatively reproduce our  experimental observations. Therefore, in our model, the incommensurability is induced by the anisotropic exchange interaction produced by the lattice distortion that was observed with x-ray diffraction measurements.~\cite{Kimura09} Because the magnetic ground state ordering  is highly sensitive to small perturbations, it is quite natural that our phase diagram differs  from those reported  in previous Refs.~\cite{Fishman_Monte2012,Haraldsen_Spin2012,Fishman11}.} We have chosen the $z$-axis along $[001]$ and the $x$ and $y$ axes along $[110]$ and $[1\bar{1}0]$, respectively  (see Fig.~\ref{f1}).

Besides the fully-polarized state,  four other spin states are stabilized in different regions of the phase diagram of $\mathcal{H}$ (see Fig.~\ref{f1}). For spatially-isotropic exchange interaction ($J'=J$) with easy-axis spin anisotropy, the  so-called ``Y" state becomes stable at low magnetic fields.~\cite{Miyashita86} The (up-up-down) UUD state, with net magnetization equal to $1/3$ of the saturation value, becomes stable above a critical field $H_c \simeq J/3$. Upon further increasing $H$, there is another transition to the so-called ``V" state that remains stable until the spins become  fully polarized. For a spatially anisotropic interaction ($J'\neq J$) without spin anisotropy, the incommensurate non-coplanar umbrella state has  lower energy than the Y phase at low fields because of its  higher uniform magnetic susceptibility.~\cite{Griset11} Besides the UUD state stabilized by thermal fluctuations at high temperatures, the umbrella state occupies most of the $H-T$ phase diagram. Here we show that the {\it{ combined}} effects of spin and spatial anisotropies in Eq.~\eqref{eq1} reproduce the measured  phase diagrams of $\mathrm{CuCrO_2}$ for both measured field orientations.

\begin{table}[b]\centering
\caption{Coplanarity $K$ and vector charity parallel (perpendicular) to the magnetic field $\chi_{\parallel}$ ($\chi_{\perp}$) in different magnetic states.}
\begin{tabular}{c | c | c | c | c | c | c }\hline\hline
    &  ICY & ICU & CY& CU & UUD & V \\
 \hline
$K$ & $=0$ & $=0$ & $>0$ & $>0$ & $=0$ & $>0$ \\
$\chi_{\parallel}$ & $=0$ & $>0$ & $=0$ & $>0$ & $=0$ & $=0$ \\
$\chi_{\perp}$ & $>0$ & $=0$ & $>0$ & $=0$ & $=0$ & $=0$ \\
 \hline \hline
\end{tabular}
\label{tbl1}
\end{table}

To discriminate between the competing spin orderings shown in Fig.~\ref{f1}, we introduce  the spin co-planarity~\cite{Watarai01}
\begin{equation}\label{eq2}
K^2=|\mathbf{K}_{12}^2|+|\mathbf{K}_{23}^2|+|\mathbf{K}_{31}^2|,
\end{equation}
where $K_{ij}=(\mathbf{m}_i\times\mathbf{m}_j)\times\mathbf{H}/H$ and $\mathbf{m}_i$ is the sublattice magnetization. Note that $K$ vanishes for incommensurate ordering because  $\mathbf{m}_i$ is parallel to ${\bf H}$ and also vanishes for the UUD phase because the moments are collinear. To discriminate between possible $K=0$ phases, we also introduce the vector chirality \cite{Griset11}
\begin{equation}\label{eq3}
\mathbf{\chi}=\frac{2}{3\sqrt{3}L^2}\sum_r\left[\mathbf{S}_{r}\times\mathbf{S}_{r+\delta_1}+\mathbf{S}_{r+\delta_1}\times\mathbf{S}_{r+\delta_2}+\mathbf{S}_{r+\delta_2}\times\mathbf{S}_{r}  \right],
\end{equation}
where $\mathbf{S}_{r}$ and $\mathbf{S}_{r+\delta_i}$ ($i=1,2$) are spins in the same triangle. We compute the components  parallel ($\chi_\parallel$) and perpendicular ($\chi_\perp$) to $\mathbf{H}$. As shown in Table~\ref{tbl1},  the combined order parameters $K$ and $\chi$ allow us to identify each of the competing spin orderings. 

A big system size is required to capture the very small deviation of $q$ from the commensurate value ($q=0.329$) that is observed in $\mathrm{CuCrO_2}$.~\cite{Soda09} Given the limitations in system sizes that are accessible for numerical simulations, we use a slightly smaller value of $q=0.3125$ (corresponding to $J'/J=0.7654$) in our MC simulations. $A_z=0.05 J$ and $A_x=0.005 J$ are typical parameters for the single-ion anisotropy terms. Details of the numerical calculation are provided in Ref.~\cite{SI}.

\begin{figure}[t]
\psfig{figure=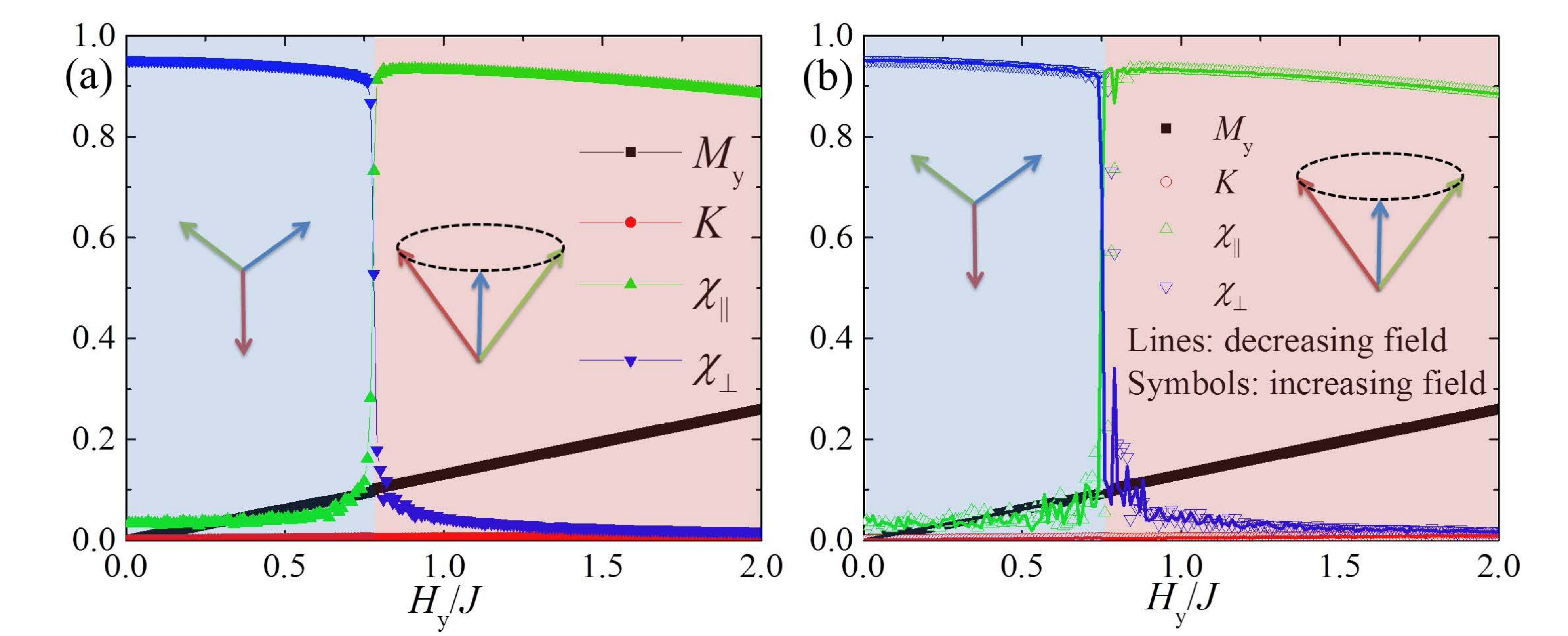,width=\columnwidth}
\caption{(color online) (a) Different observables as a function of  ${\bf H} \parallel {\hat y}$ obtained from equilibrium simulations. (b) Similar to (a) but obtained by sweeping the magnetic field. Here $T=0.02J$ and $M_y$ is the total magnetization along the field direction.} \label{f2}
\end{figure}

We first focus on the case  $\mathbf{H}\parallel \hat{y}$ shown in Fig.~\ref{f2} (a). $\chi_\parallel$ increases sharply at $H_y/J \simeq 0.8$ indicating a first-order phase transition from the incommensurate Y state to the IC umbrella state ($K=0$ indicates incommensurability). However, the magnetization curve only shows a practically unnoticeable discontinuity at the transition. This spin flop transition can be understood as follows. Small distortions of the spin configuration can be neglected  for a weak spin anisotropy. For hard-axis anisotropy along the $x$ direction, the spins in the ICY state lie in the $yz$ plane and the spin configuration for the  ICY ground state  is $\mathbf{S}=[0,\ \cos (qx),\ \sin (qx)]$. The corresponding energy is $E_Y=-A_z / 2-\chi _Y H^2/2-J- J'^2/2 J$, where $\chi_Y={J^3}{(2J + J')^{ - 2}}{(2{J^2} - 2JJ' + J{'^2})^{ - 1}}$ is the magnetic susceptibility. The spins cannot avoid the hard axis in the umbrella state: $\mathbf{S}=[\cos\theta\cos (qx),\ \cos\theta\sin (qx),\ \sin\theta]$, and its energy is $E_U=\frac{1}{2} \left(A_x-A_z\right)\cos ^2\theta  - \chi _{U} H^2/2-J-J'^2/2 J$
with $\chi_U=J{(2J + J')^{ - 2}}$ being the uniform magnetic susceptibility and $\sin\theta= H J(2J+J')^{-2}$. The Y and umbrella states have the same energy in absence of spatial and spin anisotropies regardless of the field value. Because thermal fluctuations favor collinear or coplanar states,~\cite{Shender82,Henley89} the Y state is selected at finite $T$. For $|J-J'|/J\ll 1$, the umbrella state has higher magnetic susceptibility: $\Delta\chi\equiv\chi_U-\chi_Y={(J - J')^2}/9{J^3}$. Thus, the umbrella state can be stabilized at high fields if the difference in the Zeeman energy gain outweighs the energy loss due to hard-axis anisotropy. The spin-flop transition field is estimated as $H_{f,y}=3J\sqrt{A_xJ}/|J-J'|$ when $A_x, A_z\ll 9(J-J')^2/J$. The resulting transition field is about $H_{f,y}\approx 0.9 J$, which is close to the  value of $H_{f,y}=0.81 J$ obtained from simulations [Fig.~\ref{f2}(a)]. The discrepancy arises from the fact that the spin ordering is not a pure single-${\bf q}$ state.~\cite{SI}  The spin-flop transition has to overcome the weak hard-axis anisotropy. Thus the hysteresis of this transition should also be weak. We performed additional simulations by sweeping $ H$ gradually,~\cite{SI} and the results are shown in Fig.~\ref{f2} (b). Hysteresis is absent in agreement with our experimental observations. Note also that the UUD state is absent at low $T$. The phase diagram for $\mathbf{H}\parallel \hat{y}$ is depicted in Fig.~\ref{f4} (a). The weak $T$-dependence of the transition field  is also consistent with the experiments.

\begin{figure}[t]
\psfig{figure=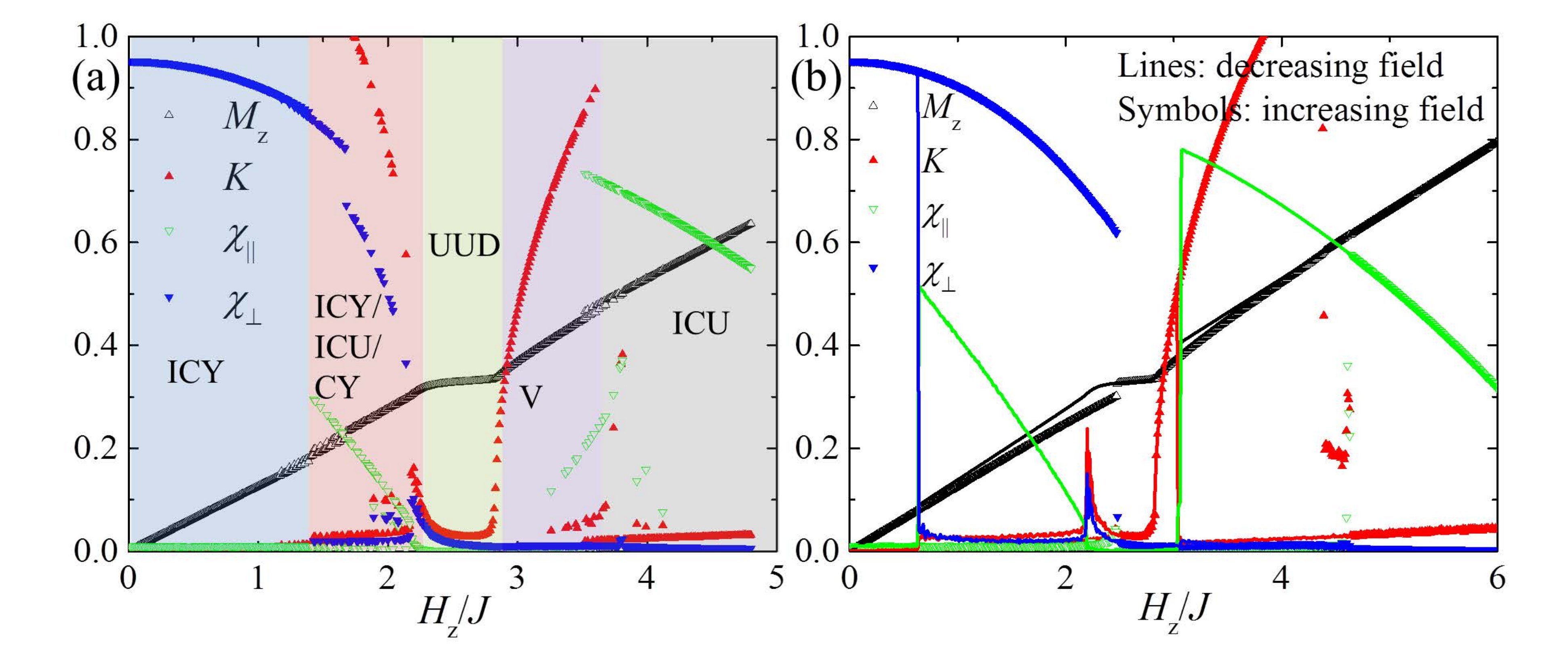,width=\columnwidth}
\caption{(color online) (a) Magnetic field dependence of different observables for $\mathbf{H}\parallel \hat{z}$ in equilibrium simulation. Due to the nature of the first order phase transition, several phases can coexist in certain magnetic field region. (b) Same as (a) but obtained with sweep of magnetic fields. Here $T=0.02J$ and $M_z$ is the total magnetization along the field direction.} \label{f3}
\end{figure}

\begin{figure}[t]
\psfig{figure=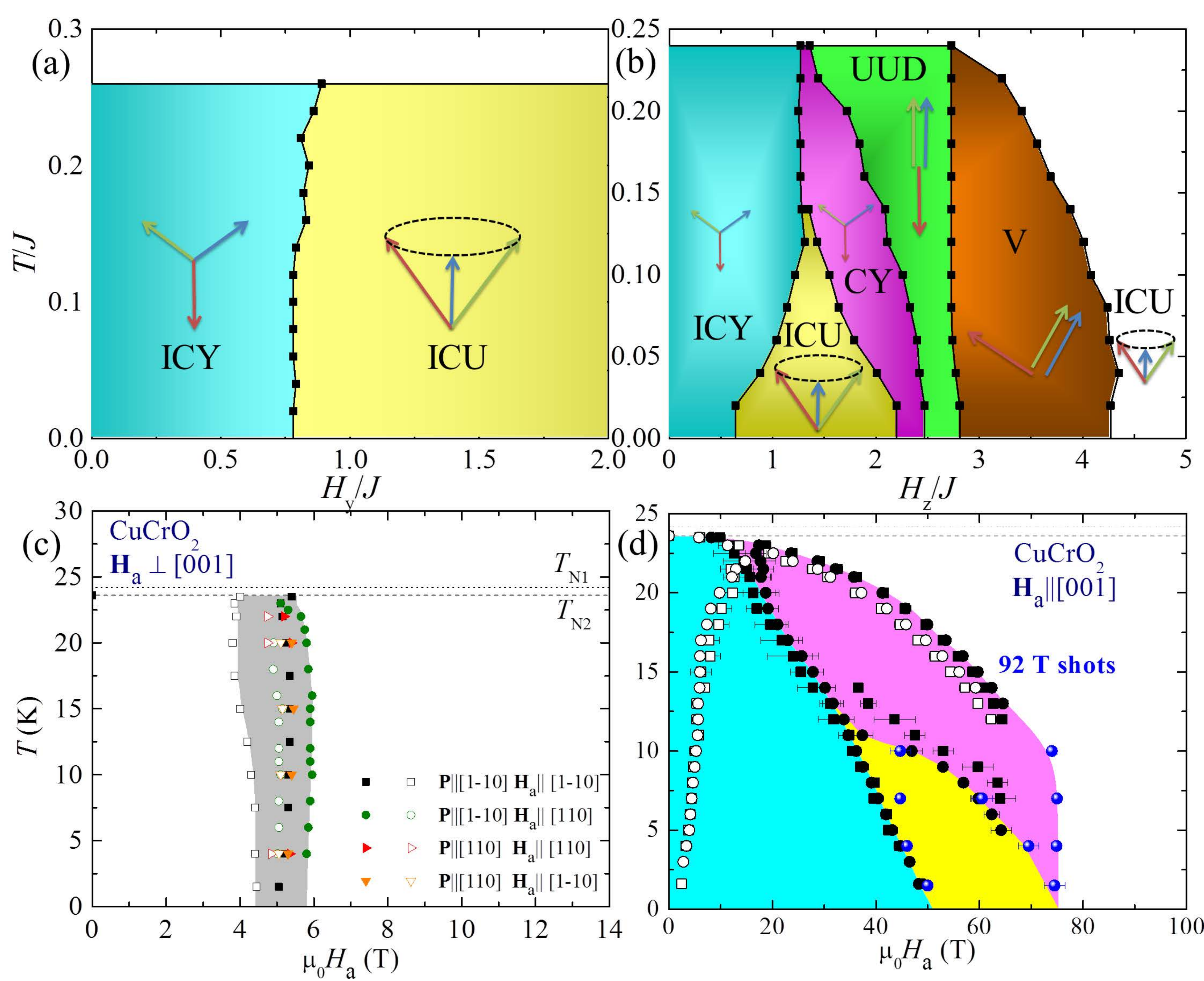,width=\columnwidth}
\caption{(color online) Calculated phase diagram of ${\cal H}$ when ${\bf H}$ is along the (a) ${\hat y}$  and (b) ${\hat z}$ directions. The phase boundary is obtained by sweeping $H$ [see Figs. \ref{f2} (b) and \ref{f3} (b)]. The phase boundaries between ICY, ICU and CY in (b) are obtained by down-sweeps of $H$.  Experimental phase diagram for (c) $\mathbf{H}_a || \hat{y}$ and (d) $\mathbf{H}_a || \hat{z}$ based on \cite{Mun13}. Solid circles and squares are based on ${\bf P} \parallel [110]$ and ${\bf P} \parallel [1\bar{1}0]$ measurements, respectively. Open and closed symbols represent transitions observed in down and
upsweeps of $H_a$, respectively. New points taken with 92 T shots are shown in blue. Phase transitions are determined from inflection points (positive or negative peaks in $dP/dt$ data), and the error bars indicate the 90\% height of the peaks.} \label{f4}
\end{figure}

Next we describe the case  ${\bf H} \parallel {\hat z}$. The results for the order parameters and magnetization are displayed in Fig.~\ref{f3}(a). Several phase transitions are observed as a function of $H_z$. The low-field  ICY state undergoes a transition into the ICU state. A second transition into the CY state occurs  before reaching the UUD state. The V state is stabilized immediately above the UUD plateau. Finally, the ICU state reappears at higher fields and remains stable until the spins become fully saturated. Except for the transition from the CY to UUD state, the transitions are strongly first order, according to the hysteresis in magnetic-sweep simulations. The UUD phase exists even at low temperatures because $H$ is now parallel to the easy-axis. The first transition from the ICY to ICU state can again be understood from simple energetic considerations.  The energy of the Y state is the same as for ${\bf H} \parallel {\hat y}$. The energy of the umbrella phase is ${E_U} = {\cos ^2}\theta {A_x} /2- A_z \sin ^2{\theta}  - \chi _U H^2 /2 - J - J{'^2}/2J.$
The energy cost of the ICU phase arises from the single-ion anisotropy. The transition field is $H_{f,z}=3J\sqrt{(A_x+A_z)J}/|J-J'|$ when $A_x, A_z\ll 9(J-J')^2/J$, which is higher than the value obtained for $\mathbf{H}\parallel \hat{y}$. Moreover, the transition requires overcoming the energy barrier $A_x+A_z$, which is bigger than the value obtained for ${\bf H} \parallel {\hat y}$. Thus, the spin flop transition for ${\bf H} || {\hat z}$ has a large hysteresis, as is clearly seen when $H$ is increased or decreased continuously [Fig. \ref{f3}(b)]. Upon increasing $H$, the system jumps from the low field ICY phase directly into the UUD plateau. Upon leaving the plateau, the spin ordering evolves into the V state and finally into the ICU state at $H_z\approx 4.5 J$. In contrast, the following sequence of phases is observed with decreasing field: ICU, V, UUD, CY, ICU, ICY. The existence of ICY and CY phases is further supported by the spin structure factor.~\cite{SI}

The precise location of the strong first-order phase transitions is difficult to determine with MC simulations.  Fig.~\ref{f4}(b) shows a rough phase diagram obtained for ${\bf H} \parallel {\hat z}$ based on the $H$-dependence of the order parameters  at different temperatures. The UUD phase appears at about 100 T for parameters  $J=2.3$ meV and $g\approx 2$ relevant to $\mathrm{CuCrO_2}$.~\cite{Poienar10}
Our simulations produce the qualitative features of the measured phase diagram,~\cite{Mun13} and we can assign the following phases as a function of increasing field: ICY, ICU, CY, and finally UUD. UUD becomes stable roughly at 1/3 of the saturation field. The magnetic states cannot be directly obtained from electric polarization measurements.~\cite{Mun13} Therefore, the proposed states can be checked by other techniques, such as muon spin spectroscopy. We remark that the phase diagram for $\mathrm{CuCrO_2}$ is very similar to another TLA compound $\mathrm{Cs_2CuCl_4}$.~\cite{Tokiwa_Magnetic2006}

\begin{figure}[t]
\psfig{figure=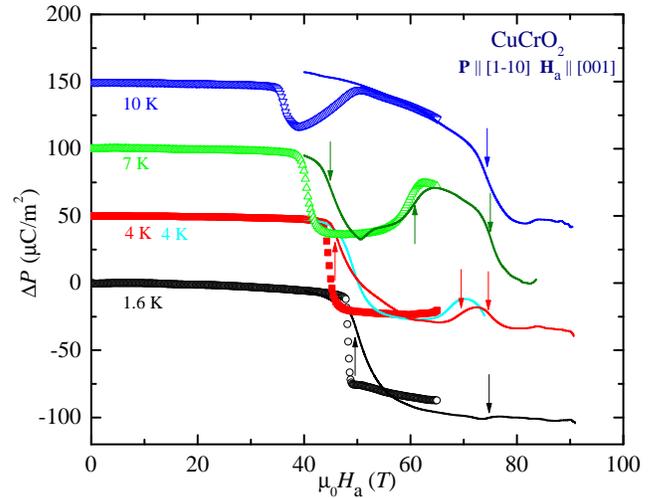,width=\columnwidth}
\caption
{(color online) ${\bf P}$ vs ${\bf H}_a$ data measured on the upsweep of a 65 T capacitor-driven magnet (points)~\cite{Mun13} and with shots up to 92 T in the 100 T multi-shot magnet (lines, this work). The variation with $H_a$ sweep rate is discussed in the text and the Supplemental Information.~\cite{SI} Arrows indicate phase transitions based on the 92 T data shown as blue points in Fig.~\ref{f4}}
\label{f5}
\end{figure}

% % % % % % experiments % % % % % % % % % % % % %
In addition to our simulations, we have extend measurements of ${\bf P}({\bf H}_a)$ to 92 T in the 100 T multi-shot magnet of the NHMFL-PFF in Los Alamos. ${\bf P} || {\hat y}$ was measured with ${\bf H}_a || {\hat x}$ and ${\hat z}$ with the same methods and samples as Mun et al. \cite{Mun13}. A poling electric field of 650 kV/m was used. For ${\bf H}_a \parallel {\hat x}$ we find no additional features in ${\bf P}({\bf H}_a)$ up to 92 Tesla (not shown), indicating that the cycloidal spiral phase persists beyond that field. The ${\bf P}$ data with ${\bf H}_a \parallel {\hat z}$ are shown in Fig.~\ref{f5} for 1.6 K $\leq T \leq$ 10 K and for upsweeps of $H_a$. Data in a 65 T magnet from \cite{Mun13} are shown for comparison. In qualitative agreement with our calculations, we observe significant differences in the width and position of the phase boundaries for different magnetic field sweep rates $dH_a/dt$. For example, at the 50 T transition, $dH_a/dt$ for a 92 T shot in the 100 T magnet is almost three times higher than for a 65 T shot in the 65 T magnet, and $dH_a/dt$ varies with maximum field in a given magnet. Sweep-rate dependences were also previously observed at the 5.3 T spin flop transition for ${\bf H}_a \perp {\hat c}$.~\cite{Mun13} With that caveat, we determine the transitions in the 92 T $P(H_a)$ data from peaks in $dP/dt$ \cite{Mun13}, and the error bars from the width of the peaks at 90\% of their height. The transitions are indicated as arrows in the $P(H_a)$ data in Fig.~\ref{f5} and as blue points in the phase diagram of Fig. \ref{f4}.

To calculate the precise evolution of ${\bf P}$ within the Arima model \cite{Arima07}, it is necessary to know the position of the O atoms. Without knowing these positions, we still expect that the electric polarization should be similar for the CY and ICY phases  because the absolute value of $q$ only changes by a very small amount. In contrast,  the  intermediate ICU phase (cycloidal spiral) should produce a rather different value of ${\bf P}$ because the magneto-electric coupling has a different origin.~\cite{Mostovoy06} This result is consistent with our measured $H_a$-dependence of ${\bf P}$ shown in Fig.~\ref{f5}.  

To summarize, we find qualitative agreement between theory and experiment. Our simple 2D model reproduces the incommensurate proper-screw spiral observed in experiments,~\cite{Soda09,Soda10} and the phase transition to an incommensurate cycloidal spiral observed for  ${\bf H}_a \parallel ab$. It also predicts a series of commensurate and incommensurate phases with increasing $H_z$, which is in rough agreement with the oscillations observed in the electric polarization. Both calculations and experiments show very strong hysteresis between up and down sweeps. Finally, our results demonstrate how a subtle competition between  spatial and spin anisotropy, magnetic frustration and  thermal fluctuations can lead to large changes of magneto-electric  properties induced by relatively small energy scales.

The authors are grateful to Yoshitomo Kamiya and Gia-Wei Chern for helpful discussions. Computer resources  were supported by the Institutional Computing Program in LANL. This work was carried out under the auspices of the NNSA of the U.S. DOE at LANL under Award No. DEAC52-06NA25396, and was supported by the U.S. Department of Energy, Office of BES "Science at 100 Tesla" program. The NHMFL Pulsed Field Facility is funded by the US National Science Foundation through Cooperative Grant No. DMR-1157490, the State of Florida, and the US Department of Energy. The research leading to these results has also received funding from the European Community's Seventh Framework Programme (FP7/2007-2013) under Grant Agreement No. 290605 (PSIFELLOW/COFUND). 

%\bibliography{reference}

\begin{thebibliography}{31}%
\makeatletter
\providecommand \@ifxundefined [1]{%
 \@ifx{#1\undefined}
}%
\providecommand \@ifnum [1]{%
 \ifnum #1\expandafter \@firstoftwo
 \else \expandafter \@secondoftwo
 \fi
}%
\providecommand \@ifx [1]{%
 \ifx #1\expandafter \@firstoftwo
 \else \expandafter \@secondoftwo
 \fi
}%
\providecommand \natexlab [1]{#1}%
\providecommand \enquote  [1]{``#1''}%
\providecommand \bibnamefont  [1]{#1}%
\providecommand \bibfnamefont [1]{#1}%
\providecommand \citenamefont [1]{#1}%
\providecommand \href@noop [0]{\@secondoftwo}%
\providecommand \href [0]{\begingroup \@sanitize@url \@href}%
\providecommand \@href[1]{\@@startlink{#1}\@@href}%
\providecommand \@@href[1]{\endgroup#1\@@endlink}%
\providecommand \@sanitize@url [0]{\catcode `\\12\catcode `\$12\catcode
  `\&12\catcode `\#12\catcode `\^12\catcode `\_12\catcode `\%12\relax}%
\providecommand \@@startlink[1]{}%
\providecommand \@@endlink[0]{}%
\providecommand \url  [0]{\begingroup\@sanitize@url \@url }%
\providecommand \@url [1]{\endgroup\@href {#1}{\urlprefix }}%
\providecommand \urlprefix  [0]{URL }%
\providecommand \Eprint [0]{\href }%
\providecommand \doibase [0]{http://dx.doi.org/}%
\providecommand \selectlanguage [0]{\@gobble}%
\providecommand \bibinfo  [0]{\@secondoftwo}%
\providecommand \bibfield  [0]{\@secondoftwo}%
\providecommand \translation [1]{[#1]}%
\providecommand \BibitemOpen [0]{}%
\providecommand \bibitemStop [0]{}%
\providecommand \bibitemNoStop [0]{.\EOS\space}%
\providecommand \EOS [0]{\spacefactor3000\relax}%
\providecommand \BibitemShut  [1]{\csname bibitem#1\endcsname}%
\let\auto@bib@innerbib\@empty
%</preamble>
\bibitem [{\citenamefont {Kadowaki}\ \emph {et~al.}(1990)\citenamefont
  {Kadowaki}, \citenamefont {Kikuchi},\ and\ \citenamefont
  {Ajiro}}]{Kadowaki90}%
  \BibitemOpen
  \bibfield  {author} {\bibinfo {author} {\bibfnamefont {H.}~\bibnamefont
  {Kadowaki}}, \bibinfo {author} {\bibfnamefont {H.}~\bibnamefont {Kikuchi}}, \
  and\ \bibinfo {author} {\bibfnamefont {Y.}~\bibnamefont {Ajiro}},\ }\href
  {http://stacks.iop.org/0953-8984/2/i=19/a=014} {\bibfield  {journal}
  {\bibinfo  {journal} {J. Phys.: Conden. Matter}\ }\textbf {\bibinfo {volume}
  {2}},\ \bibinfo {pages} {4485} (\bibinfo {year} {1990})}\BibitemShut
  {NoStop}%
\bibitem [{\citenamefont {Crottaz}\ \emph {et~al.}(1996)\citenamefont
  {Crottaz}, \citenamefont {Kubel},\ and\ \citenamefont
  {Schmid}}]{Crottaz1996}%
  \BibitemOpen
  \bibfield  {author} {\bibinfo {author} {\bibfnamefont {O.}~\bibnamefont
  {Crottaz}}, \bibinfo {author} {\bibfnamefont {F.}~\bibnamefont {Kubel}}, \
  and\ \bibinfo {author} {\bibfnamefont {H.}~\bibnamefont {Schmid}},\ }\href
  {\doibase http://dx.doi.org/10.1006/jssc.1996.0109} {\bibfield  {journal}
  {\bibinfo  {journal} {Journal of Solid State Chemistry}\ }\textbf {\bibinfo
  {volume} {122}},\ \bibinfo {pages} {247 } (\bibinfo {year}
  {1996})}\BibitemShut {NoStop}%
\bibitem [{\citenamefont {Poienar}\ \emph {et~al.}(2010)\citenamefont
  {Poienar}, \citenamefont {Damay}, \citenamefont {Martin}, \citenamefont
  {Robert},\ and\ \citenamefont {Petit}}]{Poienar10}%
  \BibitemOpen
  \bibfield  {author} {\bibinfo {author} {\bibfnamefont {M.}~\bibnamefont
  {Poienar}}, \bibinfo {author} {\bibfnamefont {F.}~\bibnamefont {Damay}},
  \bibinfo {author} {\bibfnamefont {C.}~\bibnamefont {Martin}}, \bibinfo
  {author} {\bibfnamefont {J.}~\bibnamefont {Robert}}, \ and\ \bibinfo {author}
  {\bibfnamefont {S.}~\bibnamefont {Petit}},\ }\href {\doibase
  10.1103/PhysRevB.81.104411} {\bibfield  {journal} {\bibinfo  {journal} {Phys.
  Rev. B}\ }\textbf {\bibinfo {volume} {81}},\ \bibinfo {pages} {104411}
  (\bibinfo {year} {2010})}\BibitemShut {NoStop}%
\bibitem [{\citenamefont {Frontzek}\ \emph {et~al.}(2011)\citenamefont
  {Frontzek}, \citenamefont {Haraldsen}, \citenamefont {Podlesnyak},
  \citenamefont {Matsuda}, \citenamefont {Christianson}, \citenamefont
  {Fishman}, \citenamefont {Sefat}, \citenamefont {Qiu}, \citenamefont
  {Copley}, \citenamefont {Barilo}, \citenamefont {Shiryaev},\ and\
  \citenamefont {Ehlers}}]{Frontzek11}%
  \BibitemOpen
  \bibfield  {author} {\bibinfo {author} {\bibfnamefont {M.}~\bibnamefont
  {Frontzek}}, \bibinfo {author} {\bibfnamefont {J.~T.}\ \bibnamefont
  {Haraldsen}}, \bibinfo {author} {\bibfnamefont {A.}~\bibnamefont
  {Podlesnyak}}, \bibinfo {author} {\bibfnamefont {M.}~\bibnamefont {Matsuda}},
  \bibinfo {author} {\bibfnamefont {A.~D.}\ \bibnamefont {Christianson}},
  \bibinfo {author} {\bibfnamefont {R.~S.}\ \bibnamefont {Fishman}}, \bibinfo
  {author} {\bibfnamefont {A.~S.}\ \bibnamefont {Sefat}}, \bibinfo {author}
  {\bibfnamefont {Y.}~\bibnamefont {Qiu}}, \bibinfo {author} {\bibfnamefont
  {J.~R.~D.}\ \bibnamefont {Copley}}, \bibinfo {author} {\bibfnamefont
  {S.}~\bibnamefont {Barilo}}, \bibinfo {author} {\bibfnamefont {S.~V.}\
  \bibnamefont {Shiryaev}}, \ and\ \bibinfo {author} {\bibfnamefont
  {G.}~\bibnamefont {Ehlers}},\ }\href {\doibase 10.1103/PhysRevB.84.094448}
  {\bibfield  {journal} {\bibinfo  {journal} {Phys. Rev. B}\ }\textbf {\bibinfo
  {volume} {84}},\ \bibinfo {pages} {094448} (\bibinfo {year}
  {2011})}\BibitemShut {NoStop}%
\bibitem [{\citenamefont {Vasiliev}\ \emph {et~al.}(2013)\citenamefont
  {Vasiliev}, \citenamefont {Prozorova}, \citenamefont {Svistov}, \citenamefont
  {Tsurkan}, \citenamefont {Dziom}, \citenamefont {Shuvaev}, \citenamefont
  {Pimenov},\ and\ \citenamefont {Pimenov}}]{Vasiliev14}%
  \BibitemOpen
  \bibfield  {author} {\bibinfo {author} {\bibfnamefont {A.~M.}\ \bibnamefont
  {Vasiliev}}, \bibinfo {author} {\bibfnamefont {L.~A.}\ \bibnamefont
  {Prozorova}}, \bibinfo {author} {\bibfnamefont {L.~E.}\ \bibnamefont
  {Svistov}}, \bibinfo {author} {\bibfnamefont {V.}~\bibnamefont {Tsurkan}},
  \bibinfo {author} {\bibfnamefont {V.}~\bibnamefont {Dziom}}, \bibinfo
  {author} {\bibfnamefont {A.}~\bibnamefont {Shuvaev}}, \bibinfo {author}
  {\bibfnamefont {A.}~\bibnamefont {Pimenov}}, \ and\ \bibinfo {author}
  {\bibfnamefont {A.}~\bibnamefont {Pimenov}},\ }\href {\doibase
  10.1103/PhysRevB.88.144403} {\bibfield  {journal} {\bibinfo  {journal} {Phys.
  Rev. B}\ }\textbf {\bibinfo {volume} {88}},\ \bibinfo {pages} {144403}
  (\bibinfo {year} {2013})}\BibitemShut {NoStop}%
\bibitem [{\citenamefont {Soda}\ \emph {et~al.}(2009)\citenamefont {Soda},
  \citenamefont {Kimura}, \citenamefont {Kimura}, \citenamefont {Matsuura},\
  and\ \citenamefont {Hirota}}]{Soda09}%
  \BibitemOpen
  \bibfield  {author} {\bibinfo {author} {\bibfnamefont {M.}~\bibnamefont
  {Soda}}, \bibinfo {author} {\bibfnamefont {K.}~\bibnamefont {Kimura}},
  \bibinfo {author} {\bibfnamefont {T.}~\bibnamefont {Kimura}}, \bibinfo
  {author} {\bibfnamefont {M.}~\bibnamefont {Matsuura}}, \ and\ \bibinfo
  {author} {\bibfnamefont {K.}~\bibnamefont {Hirota}},\ }\href {\doibase
  10.1143/JPSJ.78.124703} {\bibfield  {journal} {\bibinfo  {journal} {Journal
  of the Physical Society of Japan}\ }\textbf {\bibinfo {volume} {78}},\
  \bibinfo {pages} {124703} (\bibinfo {year} {2009})}\BibitemShut {NoStop}%
\bibitem [{\citenamefont {Soda}\ \emph {et~al.}(2010)\citenamefont {Soda},
  \citenamefont {Kimura}, \citenamefont {Kimura},\ and\ \citenamefont
  {Hirota}}]{Soda10}%
  \BibitemOpen
  \bibfield  {author} {\bibinfo {author} {\bibfnamefont {M.}~\bibnamefont
  {Soda}}, \bibinfo {author} {\bibfnamefont {K.}~\bibnamefont {Kimura}},
  \bibinfo {author} {\bibfnamefont {T.}~\bibnamefont {Kimura}}, \ and\ \bibinfo
  {author} {\bibfnamefont {K.}~\bibnamefont {Hirota}},\ }\href {\doibase
  10.1103/PhysRevB.81.100406} {\bibfield  {journal} {\bibinfo  {journal} {Phys.
  Rev. B}\ }\textbf {\bibinfo {volume} {81}},\ \bibinfo {pages} {100406}
  (\bibinfo {year} {2010})}\BibitemShut {NoStop}%
\bibitem [{\citenamefont {Kimura}\ \emph {et~al.}(2008)\citenamefont {Kimura},
  \citenamefont {Nakamura}, \citenamefont {Ohgushi},\ and\ \citenamefont
  {Kimura}}]{Kimura08}%
  \BibitemOpen
  \bibfield  {author} {\bibinfo {author} {\bibfnamefont {K.}~\bibnamefont
  {Kimura}}, \bibinfo {author} {\bibfnamefont {H.}~\bibnamefont {Nakamura}},
  \bibinfo {author} {\bibfnamefont {K.}~\bibnamefont {Ohgushi}}, \ and\
  \bibinfo {author} {\bibfnamefont {T.}~\bibnamefont {Kimura}},\ }\href
  {\doibase 10.1103/PhysRevB.78.140401} {\bibfield  {journal} {\bibinfo
  {journal} {Phys. Rev. B}\ }\textbf {\bibinfo {volume} {78}},\ \bibinfo
  {pages} {140401} (\bibinfo {year} {2008})}\BibitemShut {NoStop}%
\bibitem [{\citenamefont {Seki}\ \emph {et~al.}(2008)\citenamefont {Seki},
  \citenamefont {Onose},\ and\ \citenamefont {Tokura}}]{Seki08}%
  \BibitemOpen
  \bibfield  {author} {\bibinfo {author} {\bibfnamefont {S.}~\bibnamefont
  {Seki}}, \bibinfo {author} {\bibfnamefont {Y.}~\bibnamefont {Onose}}, \ and\
  \bibinfo {author} {\bibfnamefont {Y.}~\bibnamefont {Tokura}},\ }\href
  {\doibase 10.1103/PhysRevLett.101.067204} {\bibfield  {journal} {\bibinfo
  {journal} {Phys. Rev. Lett.}\ }\textbf {\bibinfo {volume} {101}},\ \bibinfo
  {pages} {067204} (\bibinfo {year} {2008})}\BibitemShut {NoStop}%
\bibitem [{\citenamefont {Kimura}\ \emph
  {et~al.}(2009{\natexlab{a}})\citenamefont {Kimura}, \citenamefont {Nakamura},
  \citenamefont {Kimura}, \citenamefont {Hagiwara},\ and\ \citenamefont
  {Kimura}}]{Kimura09PRL}%
  \BibitemOpen
  \bibfield  {author} {\bibinfo {author} {\bibfnamefont {K.}~\bibnamefont
  {Kimura}}, \bibinfo {author} {\bibfnamefont {H.}~\bibnamefont {Nakamura}},
  \bibinfo {author} {\bibfnamefont {S.}~\bibnamefont {Kimura}}, \bibinfo
  {author} {\bibfnamefont {M.}~\bibnamefont {Hagiwara}}, \ and\ \bibinfo
  {author} {\bibfnamefont {T.}~\bibnamefont {Kimura}},\ }\href {\doibase
  10.1103/PhysRevLett.103.107201} {\bibfield  {journal} {\bibinfo  {journal}
  {Phys. Rev. Lett.}\ }\textbf {\bibinfo {volume} {103}},\ \bibinfo {pages}
  {107201} (\bibinfo {year} {2009}{\natexlab{a}})}\BibitemShut {NoStop}%
\bibitem [{\citenamefont {Poienar}\ \emph {et~al.}(2009)\citenamefont
  {Poienar}, \citenamefont {Damay}, \citenamefont {Martin}, \citenamefont
  {Hardy}, \citenamefont {Maignan},\ and\ \citenamefont {Andr\'e}}]{Poienar09}%
  \BibitemOpen
  \bibfield  {author} {\bibinfo {author} {\bibfnamefont {M.}~\bibnamefont
  {Poienar}}, \bibinfo {author} {\bibfnamefont {F.~m.~c.}\ \bibnamefont
  {Damay}}, \bibinfo {author} {\bibfnamefont {C.}~\bibnamefont {Martin}},
  \bibinfo {author} {\bibfnamefont {V.}~\bibnamefont {Hardy}}, \bibinfo
  {author} {\bibfnamefont {A.}~\bibnamefont {Maignan}}, \ and\ \bibinfo
  {author} {\bibfnamefont {G.}~\bibnamefont {Andr\'e}},\ }\href {\doibase
  10.1103/PhysRevB.79.014412} {\bibfield  {journal} {\bibinfo  {journal} {Phys.
  Rev. B}\ }\textbf {\bibinfo {volume} {79}},\ \bibinfo {pages} {014412}
  (\bibinfo {year} {2009})}\BibitemShut {NoStop}%
\bibitem [{\citenamefont {Kajimoto}\ \emph {et~al.}(2010)\citenamefont
  {Kajimoto}, \citenamefont {Nakajima}, \citenamefont {Ohira-Kawamura},
  \citenamefont {Inamura}, \citenamefont {Kakurai}, \citenamefont {Arai},
  \citenamefont {Hokazono}, \citenamefont {Oozono},\ and\ \citenamefont
  {Okuda}}]{Kajimoto10}%
  \BibitemOpen
  \bibfield  {author} {\bibinfo {author} {\bibfnamefont {R.}~\bibnamefont
  {Kajimoto}}, \bibinfo {author} {\bibfnamefont {K.}~\bibnamefont {Nakajima}},
  \bibinfo {author} {\bibfnamefont {S.}~\bibnamefont {Ohira-Kawamura}},
  \bibinfo {author} {\bibfnamefont {Y.}~\bibnamefont {Inamura}}, \bibinfo
  {author} {\bibfnamefont {K.}~\bibnamefont {Kakurai}}, \bibinfo {author}
  {\bibfnamefont {M.}~\bibnamefont {Arai}}, \bibinfo {author} {\bibfnamefont
  {T.}~\bibnamefont {Hokazono}}, \bibinfo {author} {\bibfnamefont
  {S.}~\bibnamefont {Oozono}}, \ and\ \bibinfo {author} {\bibfnamefont
  {T.}~\bibnamefont {Okuda}},\ }\href {\doibase 10.1143/JPSJ.79.123705}
  {\bibfield  {journal} {\bibinfo  {journal} {Journal of the Physical Society
  of Japan}\ }\textbf {\bibinfo {volume} {79}},\ \bibinfo {pages} {123705}
  (\bibinfo {year} {2010})}\BibitemShut {NoStop}%
\bibitem [{\citenamefont {hisa Arima}(2007)}]{Arima07}%
  \BibitemOpen
  \bibfield  {author} {\bibinfo {author} {\bibfnamefont {T.}~\bibnamefont {hisa
  Arima}},\ }\href {\doibase 10.1143/JPSJ.76.073702} {\bibfield  {journal}
  {\bibinfo  {journal} {Journal of the Physical Society of Japan}\ }\textbf
  {\bibinfo {volume} {76}},\ \bibinfo {pages} {073702} (\bibinfo {year}
  {2007})}\BibitemShut {NoStop}%
\bibitem [{\citenamefont {Kimura}\ \emph
  {et~al.}(2009{\natexlab{b}})\citenamefont {Kimura}, \citenamefont {Otani},
  \citenamefont {Nakamura}, \citenamefont {Wakabayashi},\ and\ \citenamefont
  {Kimura}}]{Kimura09}%
  \BibitemOpen
  \bibfield  {author} {\bibinfo {author} {\bibfnamefont {K.}~\bibnamefont
  {Kimura}}, \bibinfo {author} {\bibfnamefont {T.}~\bibnamefont {Otani}},
  \bibinfo {author} {\bibfnamefont {H.}~\bibnamefont {Nakamura}}, \bibinfo
  {author} {\bibfnamefont {Y.}~\bibnamefont {Wakabayashi}}, \ and\ \bibinfo
  {author} {\bibfnamefont {T.}~\bibnamefont {Kimura}},\ }\href {\doibase
  10.1143/JPSJ.78.113710} {\bibfield  {journal} {\bibinfo  {journal} {Journal
  of the Physical Society of Japan}\ }\textbf {\bibinfo {volume} {78}},\
  \bibinfo {pages} {113710} (\bibinfo {year} {2009}{\natexlab{b}})}\BibitemShut
  {NoStop}%
\bibitem [{\citenamefont {Aktas}\ \emph {et~al.}(2013)\citenamefont {Aktas},
  \citenamefont {Quirion}, \citenamefont {Otani},\ and\ \citenamefont
  {Kimura}}]{Aktas13}%
  \BibitemOpen
  \bibfield  {author} {\bibinfo {author} {\bibfnamefont {O.}~\bibnamefont
  {Aktas}}, \bibinfo {author} {\bibfnamefont {G.}~\bibnamefont {Quirion}},
  \bibinfo {author} {\bibfnamefont {T.}~\bibnamefont {Otani}}, \ and\ \bibinfo
  {author} {\bibfnamefont {T.}~\bibnamefont {Kimura}},\ }\href {\doibase
  10.1103/PhysRevB.88.224104} {\bibfield  {journal} {\bibinfo  {journal} {Phys.
  Rev. B}\ }\textbf {\bibinfo {volume} {88}},\ \bibinfo {pages} {224104}
  (\bibinfo {year} {2013})}\BibitemShut {NoStop}%
\bibitem [{\citenamefont {Ehlers}\ \emph {et~al.}(2013)\citenamefont {Ehlers},
  \citenamefont {Podlesnyak}, \citenamefont {Frontzek}, \citenamefont
  {Freitas}, \citenamefont {Ghivelder}, \citenamefont {Gardner}, \citenamefont
  {Shiryaev},\ and\ \citenamefont {Barilo}}]{Ehlers13}%
  \BibitemOpen
  \bibfield  {author} {\bibinfo {author} {\bibfnamefont {G.}~\bibnamefont
  {Ehlers}}, \bibinfo {author} {\bibfnamefont {A.~A.}\ \bibnamefont
  {Podlesnyak}}, \bibinfo {author} {\bibfnamefont {M.}~\bibnamefont
  {Frontzek}}, \bibinfo {author} {\bibfnamefont {R.~S.}\ \bibnamefont
  {Freitas}}, \bibinfo {author} {\bibfnamefont {L.}~\bibnamefont {Ghivelder}},
  \bibinfo {author} {\bibfnamefont {J.~S.}\ \bibnamefont {Gardner}}, \bibinfo
  {author} {\bibfnamefont {S.~V.}\ \bibnamefont {Shiryaev}}, \ and\ \bibinfo
  {author} {\bibfnamefont {S.}~\bibnamefont {Barilo}},\ }\href
  {http://stacks.iop.org/0953-8984/25/i=49/a=496009} {\bibfield  {journal}
  {\bibinfo  {journal} {Journal of Physics: Condensed Matter}\ }\textbf
  {\bibinfo {volume} {25}},\ \bibinfo {pages} {496009} (\bibinfo {year}
  {2013})}\BibitemShut {NoStop}%
\bibitem [{\citenamefont {Fishman}(2011)}]{Fishman11}%
  \BibitemOpen
  \bibfield  {author} {\bibinfo {author} {\bibfnamefont {R.~S.}\ \bibnamefont
  {Fishman}},\ }\href {http://stacks.iop.org/0953-8984/23/i=36/a=366002}
  {\bibfield  {journal} {\bibinfo  {journal} {Journal of Physics: Condensed
  Matter}\ }\textbf {\bibinfo {volume} {23}},\ \bibinfo {pages} {366002}
  (\bibinfo {year} {2011})}\BibitemShut {NoStop}%
\bibitem [{\citenamefont {{Mun}}\ \emph {et~al.}(2014)\citenamefont {{Mun}},
  \citenamefont {{Frontzek}}, \citenamefont {{Podlesnyak}}, \citenamefont
  {{Ehlers}}, \citenamefont {{Barilo}}, \citenamefont {{Shiryaev}},\ and\
  \citenamefont {{Zapf}}}]{Mun13}%
  \BibitemOpen
  \bibfield  {author} {\bibinfo {author} {\bibfnamefont {E.}~\bibnamefont
  {{Mun}}}, \bibinfo {author} {\bibfnamefont {M.}~\bibnamefont {{Frontzek}}},
  \bibinfo {author} {\bibfnamefont {A.}~\bibnamefont {{Podlesnyak}}}, \bibinfo
  {author} {\bibfnamefont {G.}~\bibnamefont {{Ehlers}}}, \bibinfo {author}
  {\bibfnamefont {S.}~\bibnamefont {{Barilo}}}, \bibinfo {author}
  {\bibfnamefont {S.~V.}\ \bibnamefont {{Shiryaev}}}, \ and\ \bibinfo {author}
  {\bibfnamefont {V.~S.}\ \bibnamefont {{Zapf}}},\ }\href@noop {} {\bibfield
  {journal} {\bibinfo  {journal} {Phys. Rev. B}\ }\textbf {\bibinfo {volume}
  {89}},\ \bibinfo {pages} {054411} (\bibinfo {year} {2014})}\BibitemShut
  {NoStop}%
\bibitem [{\citenamefont {Yamaguchi}\ \emph {et~al.}(2010)\citenamefont
  {Yamaguchi}, \citenamefont {Ohtomo}, \citenamefont {Kimura}, \citenamefont
  {Hagiwara}, \citenamefont {Kimura}, \citenamefont {Kimura}, \citenamefont
  {Okuda},\ and\ \citenamefont {Kindo}}]{Yamaguchi10}%
  \BibitemOpen
  \bibfield  {author} {\bibinfo {author} {\bibfnamefont {H.}~\bibnamefont
  {Yamaguchi}}, \bibinfo {author} {\bibfnamefont {S.}~\bibnamefont {Ohtomo}},
  \bibinfo {author} {\bibfnamefont {S.}~\bibnamefont {Kimura}}, \bibinfo
  {author} {\bibfnamefont {M.}~\bibnamefont {Hagiwara}}, \bibinfo {author}
  {\bibfnamefont {K.}~\bibnamefont {Kimura}}, \bibinfo {author} {\bibfnamefont
  {T.}~\bibnamefont {Kimura}}, \bibinfo {author} {\bibfnamefont
  {T.}~\bibnamefont {Okuda}}, \ and\ \bibinfo {author} {\bibfnamefont
  {K.}~\bibnamefont {Kindo}},\ }\href {\doibase 10.1103/PhysRevB.81.033104}
  {\bibfield  {journal} {\bibinfo  {journal} {Phys. Rev. B}\ }\textbf {\bibinfo
  {volume} {81}},\ \bibinfo {pages} {033104} (\bibinfo {year}
  {2010})}\BibitemShut {NoStop}%
\bibitem [{\citenamefont {Frontzek}\ \emph {et~al.}(2012)\citenamefont
  {Frontzek}, \citenamefont {Ehlers}, \citenamefont {Podlesnyak}, \citenamefont
  {Cao}, \citenamefont {Matsuda}, \citenamefont {Zaharko}, \citenamefont
  {Aliouane}, \citenamefont {Barilo},\ and\ \citenamefont
  {Shiryaev}}]{Frontzek12}%
  \BibitemOpen
  \bibfield  {author} {\bibinfo {author} {\bibfnamefont {M.}~\bibnamefont
  {Frontzek}}, \bibinfo {author} {\bibfnamefont {G.}~\bibnamefont {Ehlers}},
  \bibinfo {author} {\bibfnamefont {A.}~\bibnamefont {Podlesnyak}}, \bibinfo
  {author} {\bibfnamefont {H.}~\bibnamefont {Cao}}, \bibinfo {author}
  {\bibfnamefont {M.}~\bibnamefont {Matsuda}}, \bibinfo {author} {\bibfnamefont
  {O.}~\bibnamefont {Zaharko}}, \bibinfo {author} {\bibfnamefont
  {N.}~\bibnamefont {Aliouane}}, \bibinfo {author} {\bibfnamefont
  {S.}~\bibnamefont {Barilo}}, \ and\ \bibinfo {author} {\bibfnamefont {S.~V.}\
  \bibnamefont {Shiryaev}},\ }\href
  {http://stacks.iop.org/0953-8984/24/i=1/a=016004} {\bibfield  {journal}
  {\bibinfo  {journal} {Journal of Physics: Condensed Matter}\ }\textbf
  {\bibinfo {volume} {24}},\ \bibinfo {pages} {016004} (\bibinfo {year}
  {2012})}\BibitemShut {NoStop}%
\bibitem [{Note1()}]{Note1}%
  \BibitemOpen
  \bibinfo {note} {A similar Hamiltonian with next NN AFM interaction and
  next-next NN AFM interaction was proposed based on inelastic neutron
  scattering measurements.~\cite {Poienar10,Frontzek11} The parameters derived
  in Ref.~\protect \rev@citealp {Poienar10} however lead to a collinear ground
  state at zero magnetic field, which is inconsistent with experiments (see the
  supplemental information for details). In Ref.~\cite {Frontzek11}, the
  incommensurate spiral is stabilized by a ferromagnetic interlayer coupling
  because the layers are not vertically stacked along the $z$-axis. This
  Hamiltonian was used in Refs.~\cite
  {Fishman_Monte2012,Haraldsen_Spin2012,Fishman11} for $\protect \mathrm
  {CuFeO_2}$ and $\protect \mathrm {CuCrO_2}$. It is a subtle issue whether the
  incommensurability results from the inequivalent intra-layer bonds or/and
  from the weak frustrated interlayer coupling. Here we seek for a minimal 2D
  model with only anisotropic NN AFM interactions to qualitatively reproduce
  our experimental observations. Therefore, in our model, the
  incommensurability is induced by the anisotropic exchange interaction
  produced by the lattice distortion that was observed with x-ray diffraction
  measurements.~\cite {Kimura09} Because the magnetic ground state ordering is
  highly sensitive to small perturbations, it is quite natural that our phase
  diagram differs from those reported in previous Refs.~\cite
  {Fishman_Monte2012,Haraldsen_Spin2012,Fishman11}.}\BibitemShut {Stop}%
\bibitem [{\citenamefont {Miyashita}(1986)}]{Miyashita86}%
  \BibitemOpen
  \bibfield  {author} {\bibinfo {author} {\bibfnamefont {S.}~\bibnamefont
  {Miyashita}},\ }\href {\doibase 10.1143/JPSJ.55.3605} {\bibfield  {journal}
  {\bibinfo  {journal} {Journal of the Physical Society of Japan}\ }\textbf
  {\bibinfo {volume} {55}},\ \bibinfo {pages} {3605} (\bibinfo {year}
  {1986})}\BibitemShut {NoStop}%
\bibitem [{\citenamefont {Griset}\ \emph {et~al.}(2011)\citenamefont {Griset},
  \citenamefont {Head}, \citenamefont {Alicea},\ and\ \citenamefont
  {Starykh}}]{Griset11}%
  \BibitemOpen
  \bibfield  {author} {\bibinfo {author} {\bibfnamefont {C.}~\bibnamefont
  {Griset}}, \bibinfo {author} {\bibfnamefont {S.}~\bibnamefont {Head}},
  \bibinfo {author} {\bibfnamefont {J.}~\bibnamefont {Alicea}}, \ and\ \bibinfo
  {author} {\bibfnamefont {O.~A.}\ \bibnamefont {Starykh}},\ }\href {\doibase
  10.1103/PhysRevB.84.245108} {\bibfield  {journal} {\bibinfo  {journal} {Phys.
  Rev. B}\ }\textbf {\bibinfo {volume} {84}},\ \bibinfo {pages} {245108}
  (\bibinfo {year} {2011})}\BibitemShut {NoStop}%
\bibitem [{\citenamefont {Watarai}\ \emph {et~al.}(2001)\citenamefont
  {Watarai}, \citenamefont {Miyashita},\ and\ \citenamefont
  {Shiba}}]{Watarai01}%
  \BibitemOpen
  \bibfield  {author} {\bibinfo {author} {\bibfnamefont {S.}~\bibnamefont
  {Watarai}}, \bibinfo {author} {\bibfnamefont {S.}~\bibnamefont {Miyashita}},
  \ and\ \bibinfo {author} {\bibfnamefont {H.}~\bibnamefont {Shiba}},\ }\href
  {\doibase 10.1143/JPSJ.70.532} {\bibfield  {journal} {\bibinfo  {journal}
  {Journal of the Physical Society of Japan}\ }\textbf {\bibinfo {volume}
  {70}},\ \bibinfo {pages} {532} (\bibinfo {year} {2001})}\BibitemShut
  {NoStop}%
\bibitem [{SI()}]{SI}%
  \BibitemOpen
  \href@noop {} {}\bibinfo {note} {The supplemental information contains
  comparison of energy between a collinear and the proper-screw spiral phase,
  simulation details, the spin structure factor, magnetic field pulse profile
  for the 65 T and 100 T magnets, and a discussion of why magnetocaloric or
  eddy current effects cannot account for the observed sweep rate
  dependence.}\BibitemShut {Stop}%
\bibitem [{\citenamefont {Shender}(1982)}]{Shender82}%
  \BibitemOpen
  \bibfield  {author} {\bibinfo {author} {\bibfnamefont {E.~F.}\ \bibnamefont
  {Shender}},\ }\href@noop {} {\bibfield  {journal} {\bibinfo  {journal} {Sov.
  Phys. JETP}\ }\textbf {\bibinfo {volume} {56}},\ \bibinfo {pages} {178}
  (\bibinfo {year} {1982})}\BibitemShut {NoStop}%
\bibitem [{\citenamefont {Henley}(1989)}]{Henley89}%
  \BibitemOpen
  \bibfield  {author} {\bibinfo {author} {\bibfnamefont {C.~L.}\ \bibnamefont
  {Henley}},\ }\href {\doibase 10.1103/PhysRevLett.62.2056} {\bibfield
  {journal} {\bibinfo  {journal} {Phys. Rev. Lett.}\ }\textbf {\bibinfo
  {volume} {62}},\ \bibinfo {pages} {2056} (\bibinfo {year}
  {1989})}\BibitemShut {NoStop}%
\bibitem [{\citenamefont {Tokiwa}\ \emph {et~al.}(2006)\citenamefont {Tokiwa},
  \citenamefont {Radu}, \citenamefont {Coldea}, \citenamefont {Wilhelm},
  \citenamefont {Tylczynski},\ and\ \citenamefont
  {Steglich}}]{Tokiwa_Magnetic2006}%
  \BibitemOpen
  \bibfield  {author} {\bibinfo {author} {\bibfnamefont {Y.}~\bibnamefont
  {Tokiwa}}, \bibinfo {author} {\bibfnamefont {T.}~\bibnamefont {Radu}},
  \bibinfo {author} {\bibfnamefont {R.}~\bibnamefont {Coldea}}, \bibinfo
  {author} {\bibfnamefont {H.}~\bibnamefont {Wilhelm}}, \bibinfo {author}
  {\bibfnamefont {Z.}~\bibnamefont {Tylczynski}}, \ and\ \bibinfo {author}
  {\bibfnamefont {F.}~\bibnamefont {Steglich}},\ }\href {\doibase
  10.1103/PhysRevB.73.134414} {\bibfield  {journal} {\bibinfo  {journal} {Phys.
  Rev. B}\ }\textbf {\bibinfo {volume} {73}},\ \bibinfo {pages} {134414}
  (\bibinfo {year} {2006})}\BibitemShut {NoStop}%
\bibitem [{\citenamefont {Mostovoy}(2006)}]{Mostovoy06}%
  \BibitemOpen
  \bibfield  {author} {\bibinfo {author} {\bibfnamefont {M.}~\bibnamefont
  {Mostovoy}},\ }\href {\doibase 10.1103/PhysRevLett.96.067601} {\bibfield
  {journal} {\bibinfo  {journal} {Phys. Rev. Lett.}\ }\textbf {\bibinfo
  {volume} {96}},\ \bibinfo {pages} {067601} (\bibinfo {year}
  {2006})}\BibitemShut {NoStop}%
\bibitem [{\citenamefont {Fishman}\ \emph {et~al.}(2012)\citenamefont
  {Fishman}, \citenamefont {Brown},\ and\ \citenamefont
  {Haraldsen}}]{Fishman_Monte2012}%
  \BibitemOpen
  \bibfield  {author} {\bibinfo {author} {\bibfnamefont {R.~S.}\ \bibnamefont
  {Fishman}}, \bibinfo {author} {\bibfnamefont {G.}~\bibnamefont {Brown}}, \
  and\ \bibinfo {author} {\bibfnamefont {J.~T.}\ \bibnamefont {Haraldsen}},\
  }\href {\doibase 10.1103/PhysRevB.85.020405} {\bibfield  {journal} {\bibinfo
  {journal} {Phys. Rev. B}\ }\textbf {\bibinfo {volume} {85}},\ \bibinfo
  {pages} {020405} (\bibinfo {year} {2012})}\BibitemShut {NoStop}%
\bibitem [{\citenamefont {Haraldsen}\ \emph {et~al.}(2012)\citenamefont
  {Haraldsen}, \citenamefont {Fishman},\ and\ \citenamefont
  {Brown}}]{Haraldsen_Spin2012}%
  \BibitemOpen
  \bibfield  {author} {\bibinfo {author} {\bibfnamefont {J.~T.}\ \bibnamefont
  {Haraldsen}}, \bibinfo {author} {\bibfnamefont {R.~S.}\ \bibnamefont
  {Fishman}}, \ and\ \bibinfo {author} {\bibfnamefont {G.}~\bibnamefont
  {Brown}},\ }\href {\doibase 10.1103/PhysRevB.86.024412} {\bibfield  {journal}
  {\bibinfo  {journal} {Phys. Rev. B}\ }\textbf {\bibinfo {volume} {86}},\
  \bibinfo {pages} {024412} (\bibinfo {year} {2012})}\BibitemShut {NoStop}%
\end{thebibliography}

\clearpage
\section{Supplementary material}

\begin{figure}[b]
\psfig{figure=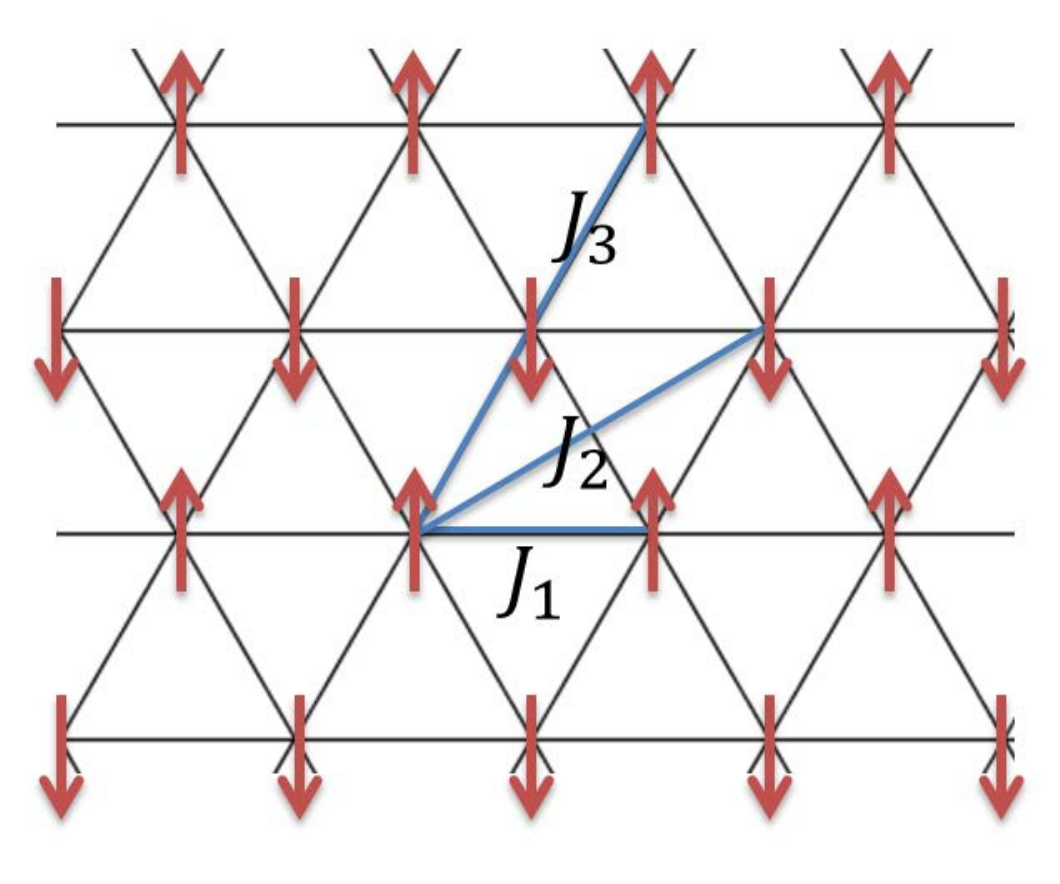,width=\columnwidth}
\caption{(color online) Schematic view of the collinear spin configuration that is stabilized at $T=0$ and ${\bf H}=0$ for the Hamiltonian parameters of Ref.~\onlinecite{Frontzek11}. Here the spins are parallel to the $z$ axis.}
\label{fS0}
\end{figure}

\section{Comparison of the energy for the collinear and proper-screw spiral phase}
Our simulations show that the parameters derived in Ref.~\onlinecite{Frontzek11} lead to the collinear spin  configuration depicted in Fig.~\ref{fS0} for $T=0$ and ${\bf H}=0$. This result can also be obtained analytically by considering the Hamiltonian used in Ref.~\onlinecite{Frontzek11}:
\begin{equation}\label{eqs0}
\mathcal{H}=\sum_{<ij>}J_{{ij}} \mathbf{S}_i\cdot \mathbf{S}_j+\sum _i\left[\frac{1}{2} A_x S_{i,x}^2-\frac{1}{2} A_z S_{i,z}^2-\mathbf{H}\cdot \mathbf{S}_i\right],
\end{equation}
which include the second and third-neighbor interactions shown in Fig.~\ref{fS0}. Here $|S|=1$. Because the $x$ and $z$ directions correspond to the hard and easy-axis, respectively, spins are parallel to the $yz$ plane for the spiral phase and to  the $z$-axis for the collinear phase. The weak interlayer coupling is neglected for the moment. For $T=0$ and ${\bf H}=0$, the energy per site of the $120^\circ$ proper-screw phase  is
\begin{equation}\label{eqs1}
E_s=-\frac{3}{2}J_1+3J_2-\frac{3}{2} J_3-\frac{1}{4} A_z,
\end{equation}
while the energy for the collinear configuration in Fig.~\ref{fS0} is
\begin{equation}
E_c=-J_1-J_2+3J_3-\frac{1}{2}A_z.
\end{equation}
For the Hamiltonian parameters of Ref.~\onlinecite{Frontzek11}, $J_1=2.8$ meV, $J_2=0.48$ meV, $J_3=0.08$ meV and $A_z=0.96$ meV, we have $E_s=-3.12$ meV and $E_c=-3.52$ meV. The weak interlayer coupling, $|J_z|=0.02$ meV, is clearly not sufficient to stabilize the proper-screw phase. Therefore,  the parameters provided by Ref.~\onlinecite{Frontzek11} lead to the  collinear spin state depicted in Fig.~\ref{fS0}, which is inconsistent with the experimental observations.

\section{Simulation details}
In our simulations, we first anneal the systems from a high temperature to a target temperature using the standard Metropolis algorithm. 
The MC measurements start  after the system reaches equilibrium. Typically we use $5\times 10^5$ MC Sweeps (MCS) for annealing, $5\times 10^5$ MC Sweeps (MCS) for equilibration, and another $5\times 10^5$ MCS for measurements. The typical system size is $L\times L=48\times 48$, but  larger sizes of $L=96$ were used to verify the irrelevance of finite size effects. 

We have also performed additional simulations by sweeping magnetic fields gradually. In this case, after increasing $H$ by 0.01$J$, we  equilibrate the system with $10^4$ MCS and perform measurements for another $10^4$ MCS.

\begin{figure}[t]
\psfig{figure=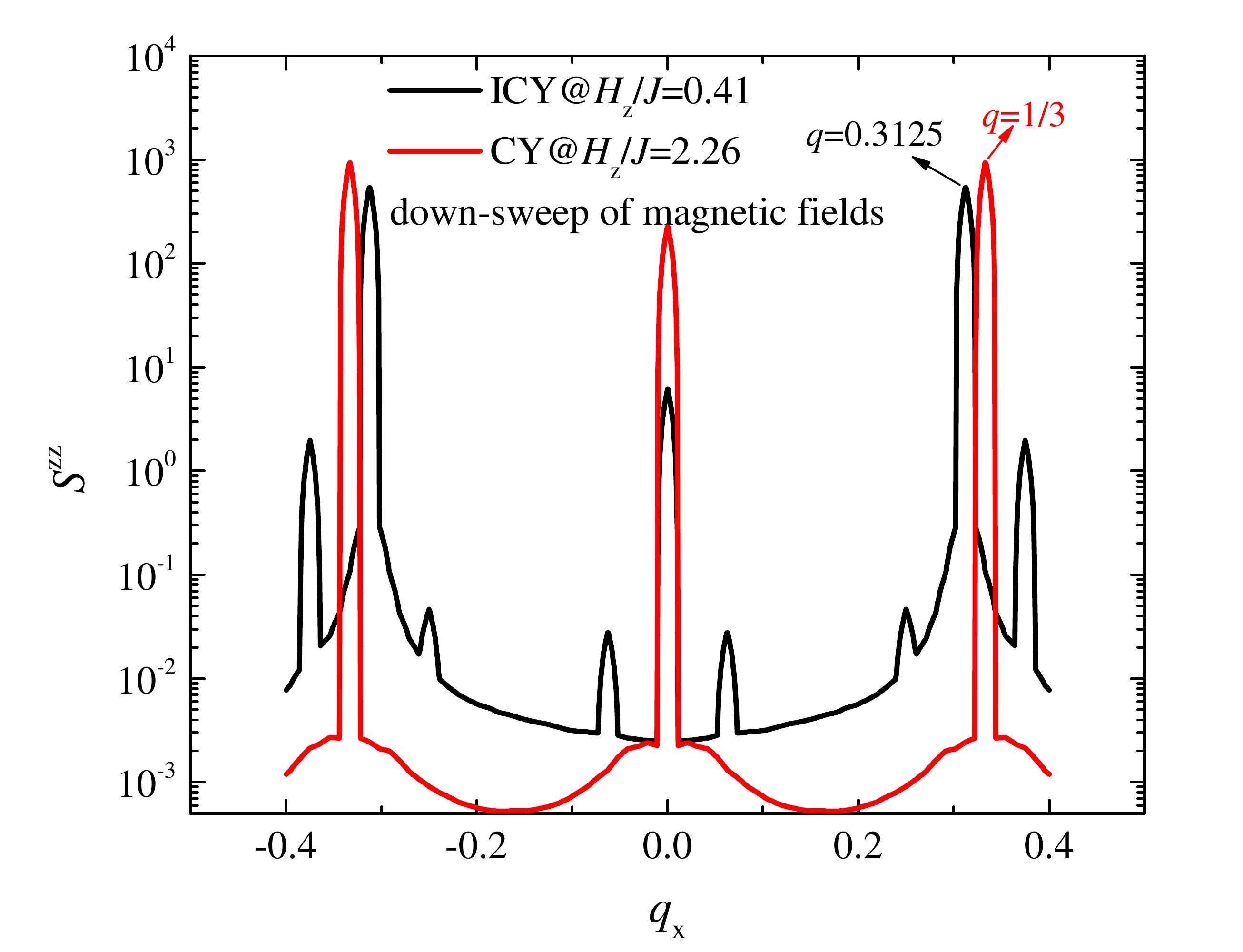,width=\columnwidth}
\caption{(color online) Spin structure factor ${S}^{zz}(q_x, q_y=0)$ obtained by down-sweep of magnetic fields  at $T=0.02 J$.}
\label{fS1}
\end{figure}

\begin{figure}[b]
\psfig{figure=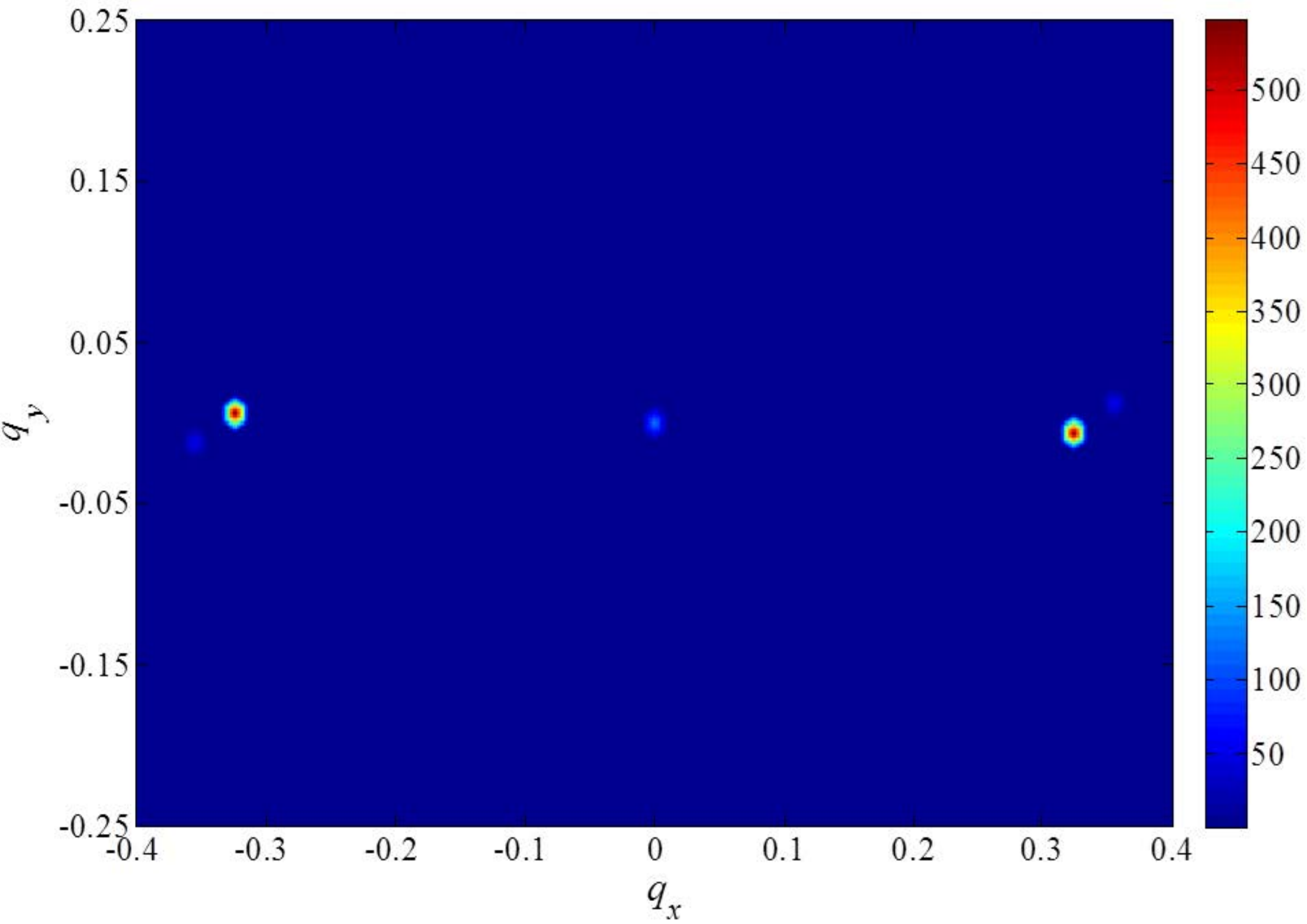,width=\columnwidth}
\caption{(color online) Spin structure factor ${S}^{zz}(\mathbf{q})$ near the ICY-ICU phase boundary at $T=0.02 J$.}
\label{fS2}
\end{figure}
\section{Spin structure factor}
To characterize the transition from the incommensurate Y (ICY) phase to commensurate Y (CY) phase, we have also calculated the spin structure factor
\begin{equation}
S^{\mu \mu} ({\bf q})= \frac{1}{N} \sum_{i,j} \exp[{i {\bf q} \cdot {({\bf r}_i - {\bf r}_j )}}] S^{\mu}_{\bf r_i} S^{\mu}_{\bf r_j},
\end{equation} 
where $\mu=x,\ y,\ z$. The calculations of ${S}^{zz}(\mathbf{q})$ for the CY and ICY phases obtained by down-sweep of the magnetic fields are shown in Fig.~\ref{fS1}. It is clear that the optimal $q$ value deviates from $q=1/3$ for the CY phase to $q=0.3125$ for the ICY phase.  ${S}^{zz}(\mathbf{q})$ has a peak at $q=0$ because there is a uniform $S_z$ component induced by $H_z$. The peak broadening of ${S}^{zz}(\mathbf{q})$ in Fig.~\ref{fS1} is mainly due to finite size effects: $\Delta q = 2 \pi /N$ (the thermal broadening is much smaller at $T=0.02 J$). 

Besides the main peak at the optimal $q$, there are additional secondary peaks shown in Fig.~\ref{fS1}. These secondary peaks are in general smaller than the main peak by several orders of magnitude. However, the amplitude of the secondary peaks increases and becomes only  one order of magnitude smaller than the amplitude of the main peak near the ICY-ICU phase boundary (see Fig. \ref{fS2}). This observation explains the small discrepancy between the transition field derived analytically for single-${\bf q}$ orderings and the value that is obtained from our MC simulations.

\begin{figure}[t]
\psfig{figure=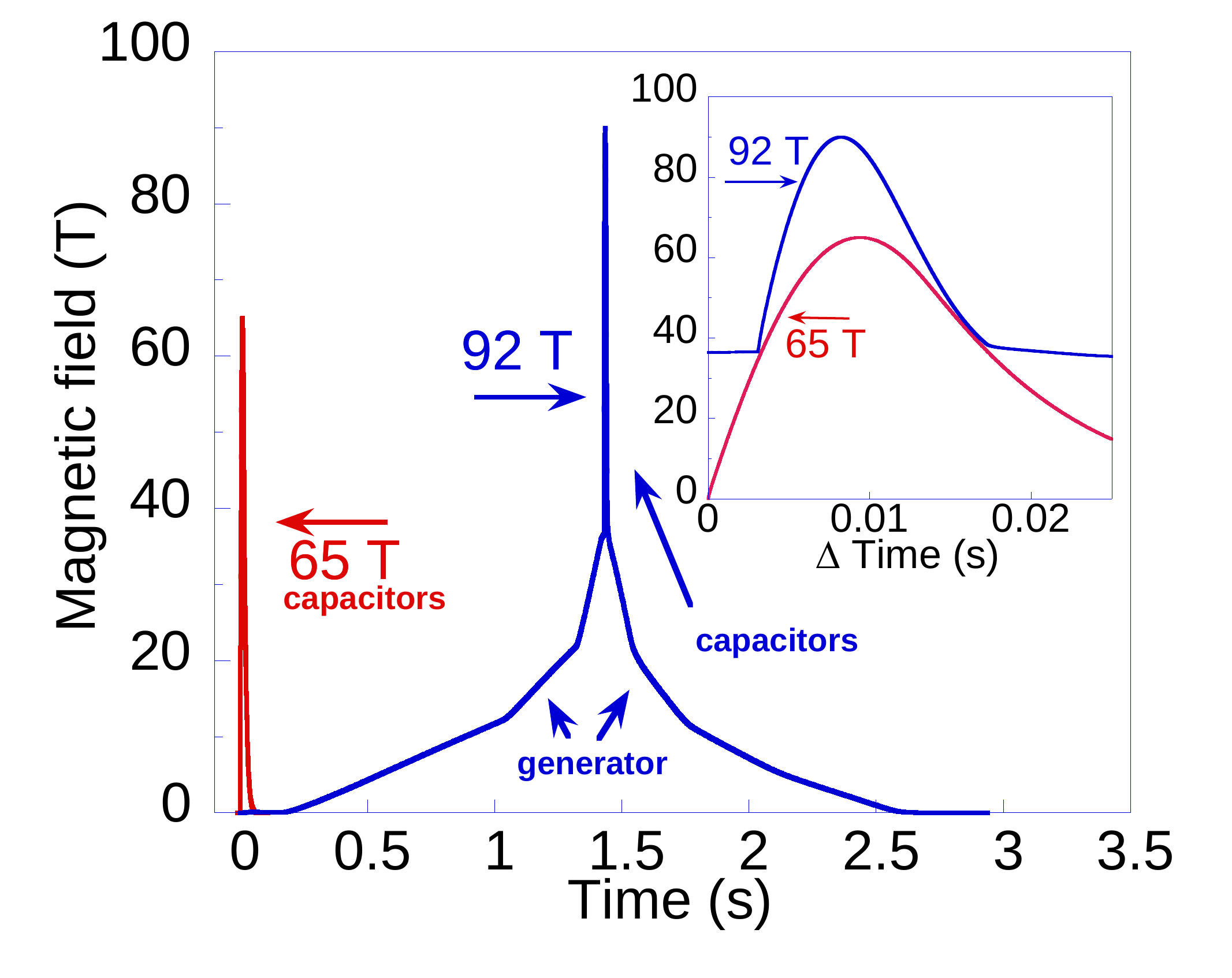,width=\columnwidth}
\caption{(color online) Magnetic field vs time profile of a 65 T shot in the 65 T capacitor-driven magnet (red), and a 92 T shot in the 100 T generator and capacitor-driven magnet (blue). The inset shows the same data shown zoomed to the high-field capacitor-driven region.}
\label{fS3}
\end{figure}

\section{Experimental details}
Figure~\ref{fS3} shows the magnetic field $H_a$ versus time $t$ for the 65 T capacitor-driven magnet, and for the 100 T magnet that is driven by a combination of a three-phase generator and a capacitor bank. The slow features are due to the generator, and the sharp spike due to the capacitor bank. At 50 T, $dH_a/dt$ is 7.4 T/s for a 65 T pulse in the capacitor-driven 65 T magnet, and $dH_a/dt$ is 20 T/s for a 92 T pulse in the capacitor-and-generator-driven 100 T magnet. In the 65 T magnet, $dH_a/dt$ at a given field scales with the peak field of the magnetic field pulse. In the 100 T magnet the same relation holds only for the capacitor-driven portion of the pulse. 

In our experimental data we can rule out significant sweep-rate dependences created by eddy currents or magnetocaloric heating for the following reasons: 1) with increasing $dH_a/dt$, the $\sim 50$ T transition shifts to higher magnetic fields, while with increasing temperature this transition moves to lower magnetic fields. Furthermore, no dramatic differences in transition widths are seen as a function of temperature, and no differences are observed between the $< 4$ K data where the sample is cooled by immersion in liquid helium, and the $> 4$ K data where the sample is  less efficiently cooled by helium gas. On the other hand, the width of the transition changes dramatically with $dH_a/dt$.

\end{document}